\documentclass[10pt,twocolumn,secnumarabic, nobibnotes, aps, prd,superscriptaddress]{revtex4-2}
\usepackage{xcolor,soul}
\usepackage{amsmath}
\usepackage{wrapfig}
\usepackage{lipsum}
\usepackage{subfigure}
\usepackage{wrapfig}
\usepackage{xparse,xcoffins}
\usepackage{graphicx}
\usepackage{comment}
\sethlcolor{red}

\ExplSyntaxOn
\NewCoffin\imagecoffin
\NewCoffin\labelcoffin
\keys_define:nn { miguel/label }
 {
  label   .tl_set:N = \l_miguel_label_tl,
  labelbox .bool_set:N = \l_miguel_label_box_bool,
  labelbox .default:n = true,
  fontsize .tl_set:N = \l_miguel_label_size_tl,
  fontsize .initial:n = \footnotesize,
  pos .choice:,
  pos/nw .code:n = \tl_set:Nn \l_miguel_label_pos_tl { left,up },
  pos/ne .code:n = \tl_set:Nn \l_miguel_label_pos_tl { right,up },
  pos/sw .code:n = \tl_set:Nn \l_miguel_label_pos_tl { left,down },
  pos/se .code:n = \tl_set:Nn \l_miguel_label_pos_tl { right,down },
  pos/n .code:n = \tl_set:Nn \l_miguel_label_pos_tl { hc,up },
  pos/w .code:n = \tl_set:Nn \l_miguel_label_pos_tl { left,vc },
  pos/s .code:n = \tl_set:Nn \l_miguel_label_pos_tl { hc,down },
  pos/e .code:n = \tl_set:Nn \l_miguel_label_pos_tl { right,vc },
  pos .initial:n = nw,
  unknown .code:n   = \clist_put_right:Nx \l_miguel_label_clist
                       { \l_keys_key_tl = \exp_not:n { #1 } }
 }
\clist_new:N \l_miguel_label_clist
\box_new:N \l_miguel_label_box
\box_new:N \l_miguel_label_image_box
\NewDocumentCommand{\xincludegraphics}{O{}m}
 {
  \group_begin:
  \tl_clear:N \l_miguel_label_tl
  \clist_clear:N \l_miguel_label_clist
  \keys_set:nn { miguel/label } { #1 }
  \tl_if_empty:NTF \l_miguel_label_tl
   {
    \miguel_includegraphics:Vn \l_miguel_label_clist { #2 }
   }
   {
    \SetHorizontalCoffin\imagecoffin
     {
      \miguel_includegraphics:Vn \l_miguel_label_clist { #2 }
     }
    \SetHorizontalCoffin\labelcoffin
     {
      \raisebox{\depth}
       {
        \bool_if:NTF \l_miguel_label_box_bool
         { \fcolorbox{white}{white}{\l_miguel_label_size_tl\l_miguel_label_tl} }
         { \l_miguel_label_size_tl\l_miguel_label_tl }
       }
     }
    \SetVerticalPole\imagecoffin{left}{3pt+\CoffinWidth\labelcoffin/2}
    \SetVerticalPole\imagecoffin{right}{\Width-3pt-\CoffinWidth\labelcoffin/2}
    \SetHorizontalPole\imagecoffin{up}{\Height-3pt-\CoffinHeight\labelcoffin/2}
    \SetHorizontalPole\imagecoffin{down}{3pt+\CoffinHeight\labelcoffin/2}
    \use:x{\JoinCoffins\imagecoffin[\l_miguel_label_pos_tl]\labelcoffin[vc,hc]}
    \TypesetCoffin\imagecoffin
   }
   \group_end:
 }
\NewDocumentCommand{\setlabel}{m}
 {
  \keys_set:nn { miguel/label } { #1 }
 }
\cs_new_protected:Nn \miguel_includegraphics:nn
 {
  \includegraphics[#1]{#2}
 }
\cs_generate_variant:Nn \miguel_includegraphics:nn { V }
\ExplSyntaxOff
\makeatletter
\newcommand{\twocolumncaption}{\@dblarg\@twocolumncaption}
\def\@twocolumncaption[#1]#2{  \renewcommand{\@makecaption}[2]{    \par\vskip\abovecaptionskip\begingroup\small\rmfamily
    \splittopskip=0pt
    \setbox\@tempboxa=\vbox{
      \@arrayparboxrestore \let \\\@normalcr
      \hsize=.5\hsize \advance\hsize-1em
      \let\\\heading@cr
      \noindent ##1\ ##2\par    }    \vbadness=10000
    \setbox\z@=\vsplit\@tempboxa to .55\ht\@tempboxa
    \setbox\z@=\vtop{\hrule height 0pt \unvbox\z@}
    \setbox\tw@=\vtop{\hrule height 0pt \unvbox\@tempboxa}
    \noindent\box\z@\hfill\box\tw@\par
    \endgroup\vskip \belowcaptionskip
  }  \setlength{\abovecaptionskip}{4ex}  \caption[#1]{#2}}

\begin{document}

\title{Vortex droplets and lattice patterns in two-dimensional traps: A
photonic spin-orbit-coupling perspective}
\author{S. Sanjay}
\email{s\textunderscore sanjay@cb.students.amrita.edu}
\affiliation{Department of Physics, Amrita School of Physical Sciences,
Amrita Vishwa Vidyapeetham, Coimbatore, 641112, Tamil Nadu, India.}

\author{S. Saravana Veni}
\email{s\textunderscore saravanaveni@cb.amrita.edu.in}
\affiliation{Department of Physics, Amrita School of Physical Sciences,
Amrita Vishwa Vidyapeetham, Coimbatore, 641112, Tamil Nadu, India.}

\author{Boris A. Malomed}
\email{malomed@tauex.tau.ac.il}

\affiliation{Department of Physical Electronics, School of Electrical Engineering, Faculty of Engineering,
Tel Aviv University, Tel Aviv 69978, Israel.}
\affiliation{Instituto de Alta Investigaci\'{o}n, Universidad de
Tarapac\'{a}, Casilla 7D, Arica, Chile}

\begin{abstract}
In the context of the mean-field exciton-polariton (EP) theory with balanced
loss and pump, we investigate the formation of lattice structures built of
individual vortex-antivortex (VAV) bound states under the action of the
two-dimensional harmonic-oscillator (HO) potential trap and effective
spin-orbit coupling (SOC), produced by the TE-TM splitting in the polariton
system. The number of VAV elements (\textquotedblleft pixels") building the
structures grow with the increase of self- and cross-interaction
coefficients. Depending upon their values and the trapping frequency, stable
ring-shaped, circular, square-shaped, rectangular, pentagonal, hexagonal, and
triangular patterns are produced, with the central site left vacant or
occupied in the lattice patterns of different types. The results suggest the experimental creation of the new patterns and their possible use for the design of integrated circuits in EP setups, controlled by the strengths of the
TE-TM splitting, nonlinearity, and HO trap.
\end{abstract}

\maketitle


\section{Introduction}

Systems of interacting quantum fluids offer broad opportunities for
investigating various macroscopic quantum phenomena, such as supersolidity
\cite{boninsegni2012colloquium,recati2023supersolidity,ilzhofer2021phase},
vorticity \cite{verhelst2017vortex, klaus2022observation, chaika2023making},
quantum droplets (QDs) \cite{Bulgac,luo2021new, guo2021new}, and others. In
particular, the studies of QDs gained impetus in low-temperature physics due
to the possibility of their experimental observation in Bose-Einstein
condensates (BECs) \cite{neely2010observation,weiler2008spontaneous}.
Recently, much interest has been drawn to this subject due to the forecast
\cite{Petrov,Astrakharchik} and synthesis of ultra-diluted QDs in binary
homonuclear \cite{cheiney2018bright,semeghini2018self} and heteronuclear
\cite{d2019observation,burchianti2020dual} Bose-Einstein condensates, where
they are sustained by the stable equilibrium between the contact
(short-range) mean-field interactions and the corrections to them, induced by
quantum fluctuations. The QD stability relies on the equilibrium between the
surface tension and bulk energy of the droplets. In addition to the creation
of stable QDs, their collisions in 1D \cite{katsimiga2023interactions}, 2D
\cite{hu2022collisional}, and 3D \cite{ferioli2019collisions} geometries
have also been studied (experimentally, in the latter case), as well as
scattering of 1D QDs on localized potentials \cite{debnath2023interaction}.

In addition to BECs with contact interactions, the bosonic condensates of
magnetic atoms with long-range dipolar interactions provide the setting for
the creation of stable anisotropic QDs and lattice patterns built of them
\cite%
{Pfau1,Pfau2,wachtler2016ground,xi2016droplet,Boronat,young2022supersolid,young2022supersolidbox}%
, including, in particular, a nearly spatially periodic stripe-shaped one,
known as the superstripe state, which is maintained by the 3D
harmonic-oscillator (HO) trapping potential \cite{young2023mini}. These
Spatially structured configurations may represent ground states and
metastable ones. Further, stable 2D \cite%
{li2018two,dong2022internal,dong2022bistable} and 3D \cite%
{kartashov2018three} vortex QDs with embedded angular momentum have been
predicted too; see a brief review in Ref. \cite{Li2024CanVQ}. On the other
hand, vortex QDs in dipolar BEC were found to be unstable \cite{Macri}. The
stability of 3D vortex QDs can also be provided by a toroidal trapping
potential \cite{LDong1}. A similar mechanism secures the stabilization of 2D
vortex QDs with multiple vorticity \cite{LDong2,LDong3} and multipole 2D QDs
\cite{LDong4}. Furthermore, the stability of the droplets can be assessed through modulation instability \cite{tabi2025coupled,veni2024numerical}. Recently \cite{tabi2023modulational} examined the modulation instability of BEC in presence of impurities. QDs were explored too in binary condensates with spin-orbit coupling (SOC) \cite{NJP-2017,cui2018spin,gangwar2024spectrum,xu2024polarized}.
Generally, the interplay of SOC with the intrinsic nonlinearity makes it possible to create various lattice-shaped and localized patterns in binary BEC \cite%
{Sakaguchi-Li,Ben-Li,meng2016experimental,zhang2016properties,zhong2018self}%
. Recent studies have also explored the existence and stability
of vortex solitons under the joint action of SOC and Rydberg
interactions \cite{wang2024rydberg,zhao2024three}. Parallel to the studies
of this phenomenology in BEC, SOC effects have also been explored in photonics.
Specifically, stable two-dimensional matter-wave solitons sustained by SOC
in binary BEC in the form of semi-vortices and mixed modes \cite{Ben-Li},
can be emulated by spatiotemporal propagation of light in a dual-core
nonlinear optical waveguide, where the effective SOC is characterized by the temporal dispersion of the inter-core coupling \cite{Konotop}.

In the realm of photonics, SOC effects play a significant role in
exciton-polariton (EP) condensates in semiconductor microcavities \cite%
{sakaguchi2017spin,whittaker2018exciton,klaas2019nonresonant,
aristov2022screening,li2022manipulating}. Polaritons are hybrid modes which
couple a material component, represented by excitons in quantum wells and cavity photons. They are represented by a two-component pseudospinor wave
function. In contrast to atomic BECs, the dynamics of EP condensates are
nonconservative, owing to significant material and defect-induced losses in
the cavities. The advantage of employing EP condensates as macroscopic
(classical) simulators of the SOC phenomena in quantum matter lies in the ease of identifying polariton states through emitted light in the host microcavity. The polarization of photons in microcavities at oblique angles gives rise to different transmission and reflection
regimes, unlike the case of the normal incidence \cite%
{dufferwiel2015spin,solnyshkov2021microcavity,lovett2023observation}. This phenomenon leads to the splitting of transverse-electric and magnetic (TE and
TM) modes, which play a fundamental role in the realization of the %
photonic SOC; it also promotes the distribution of
spin-polarized polaritons and causes periodic oscillations of the photonic
pseudospin \cite{ma2020chiral}. The effect of the polarization structure of
photons on EP states give rise to a profound nonlinearity that facilitates
the emergence of self-sustained nonlinear topological modes, such as
half-vortices \cite{cheng2024topological,pukrop2020circular}, topological
polaritons \cite{solnyshkov2021microcavity}, and skyrmions \cite%
{cilibrizzi2016half}. The nonlinearity may also be used for the development
of applications \cite{lagoudakis2017polariton,liew2018quantum}. Recent
experiments confirmed the generation of vortices in EP condensates \cite%
{gnusov2023quantum}, utilizing a cylindrically asymmetric in-plane optical
trap imposed by a composite nonresonant excitation beam. Furthermore, the
creation of half-vortices through the interaction of the TE-TM and Zeeman
splittings under the action of the ring-shaped potential, produced by an
external nonresonant depolarized pump has been demonstrated in Ref. \cite%
{yulin2020spinning}.
While the actual confinement in polariton condensates is provided by
Bragg reflectors and detuning effects, rather than an external HO trap, it
may be used as an effective confinement model \cite{tosi2012sculpting}. In
this context, the TE-TM splitting affects the stability of half-vortices in
the conservative setting \cite{toledo2010comment,flayac2010reply}, as well
as in the presence of loss and gain \cite{lobanov2010stable}. Accordingly, the half-vortex profiles are \textquotedblleft warped" by the TE-TM
splitting, losing their axial symmetry \cite{toledo2014warping}. The recent
analysis of the modulation instability under the action of the photonic SOC
\cite{madimabe2023modulational} reveals the emergence of wave modes and
opens ways for exploring novel nonlinear phenomena in EP condensates that
incorporate the TE-TM splitting and magnetic fields, such as the
condensate-reservoir regime \cite{bobrovska2015adiabatic} and multimode
dynamics \cite{bobrovska2014stability}. In addition to this, the inclusion
of the Raman effect is likely to enhance instability phenomena in the
dynamics of EP condensates \cite{tabi2024effect}.

EP interaction systems may be used in various applications, such as quantum-information processing, where polaritons can be manipulated to design quantum gates \cite{karnieli2024universal}, and EP-based lasers, that may operate more efficiently than usual lasers \cite{zhang2022recent}.

The objective of the present work is to study the dynamics of EP condensates in the context of an effectively 2D conservative system, specifically focusing on SOC induced by the TE-TM splitting. In the two-component (spinor) system of
Gross-Pitaevskii (GP) equations, which model the EP condensate, SOC is
represented by the second-order differential operator, unlike the
first-order operator which represents SOC in BEC of ultracold atomic gases.
Due to this fact, the GP system for the EP condensates gives rise to
two-component bound states of the vortex-antivortex (VAV) type \cite%
{sakaguchi2017spin}, in addition to semi-vortices (SVs), that exist as
stable states in the case of the first-order SOC operator \cite{Ben-Li}. In
this work, we demonstrate that multiple VAV elements (\textquotedblleft
pixels") can build stable circular, hexagonal, triangular, and pentagonal
lattice configurations. We address the settings which include the HO
trapping potential acting in the 2D plane $\left( x,y\right) $, while it is
implied that the reduction of the original 3D system to its 2D form is
provided by the action of the tight confinement in the transverse direction.
The stability of the VAV bound state and various lattice patterns built of
VAV pixels are established by dint of numerical simulations of the long-time
evolution.

The paper is organized as follows. In Section II, we present the nonlinear
mean-field (GP) model for the EP condensate under the action of the TE-TM
splitting and 2D trapping potential. In Section III, we present
systematically generated numerical findings for various lattice
configurations. The paper is concluded in Section IV.

\section{The model}

EP condensates trapped in microcavities are quantum fluids that exhibit
nonlinear dynamical properties. Unlike atomic BECs, polaritons are
inherently dissipative modes whose dynamics are affected by the balance
between the gain, which is provided by a reservoir, and loss, determined by
a finite EP lifetime. The dynamics of the interacting EP condensates,
confined by the 2D HO potential,%
\begin{equation}
U\left( r\right) =\left( \omega ^{2}/2\right) r^{2},~r^{2}\equiv x^{2}+y^{2},
\label{U}
\end{equation}%
in the presence of the TE-TM splitting is modeled by the coupled GP
equations for components $\psi _{1,2}$ of the EP spinor wave function \cite%
{cheng2024topological,aristov2022screening,madimabe2023modulational}:
\begin{subequations}
\label{gpe}
\begin{gather}
i\frac{\partial \psi _{1}}{\partial t}=-\bigg[\frac{1}{2}\nabla ^{2}+U(r)+%
\frac{\delta g}{2}\left( |\psi _{1}|^{2}+|\psi _{2}|^{2}\right)\nonumber  \\
+g\left( |\psi _{1}|^{2}-|\psi _{2}|^{2}\right) +\big(g_{R}n_{1}
+\bar{g_{R}}n_{2}\big)\\
+i\bigg(\frac{Rn_{1}}{2}-\gamma _{c}\bigg)\bigg]\psi
_{1}+\sigma \left( \frac{\partial }{\partial x}-i\frac{\partial }{\partial y}%
\right) ^{2}\psi _{2},\nonumber \\
\qquad i\frac{\partial \psi _{2}}{\partial t}=-\bigg[\frac{1}{2}\nabla
^{2}+U(r)+\frac{\delta g}{2}\left( |\psi _{2}|^{2}+|\psi _{1}|^{2}\right)\nonumber  \\
-g\left( |\psi _{2}|^{2}-|\psi _{1}|^{2}\right) +\big(g_{R}n_{2}
+\bar{g_{R}}n_{1}\big)\\+i\bigg(\frac{Rn_{2}}{2}-\gamma _{c}\bigg)\bigg]\psi
_{2} +\sigma \left( \frac{\partial }{\partial x}+i\frac{\partial }{\partial y}%
\right) ^{2}\psi _{1}\nonumber.
\end{gather}%
\end{subequations}
Here components $\psi _{1}$ and $\psi _{2}$ represent different polariton polarizations \cite{flayac2010topological}. $\nabla ^{2}$ denotes 2D Laplacian, the normalized trapping frequency in Eq. (\ref{U}) is $\omega \equiv 2\pi f/\omega _{\perp }$, where
$f$ is the frequency value in Hz, while $\omega _{\perp }$is the frequency
accounting for the strong transverse confinement, and $\sigma $ represents
the strength of the effective photonic SOC, i.e., the effect of the TE-TM
splitting on the EP modes \cite{aristov2022screening}, \cite%
{dovzhenko2023next}. As mentioned above, the TE-TM splitting in EP systems
leads to anisotropic dispersion and spin-dependent coupling, enriching the
system's dynamics and allowing exploration of spin-textured patterns and
polarization vortices. Further, $g$ is the intracomponent interaction
coefficient, and
\begin{equation}
\delta g\equiv g+g_{12},  \label{delta-g}
\end{equation}%
where $g_{12}$ is the coefficient accounting for the inter-component
interactions. We here address the physically relevant case of $g>0$
and $g_{12}<0$, which implies the repulsion and attraction of EPs with
identical and opposite spins, respectively. The repulsive and attractive
interactions in the EP system may be nearly canceled
by means of the Feshbach resonance \cite{takemura2014polaritonic}, which leads to $|\delta g|\ll g$.

The rate equations for reservoir densities $n_{1,2}$, coupled to the GP
system are written as
\begin{equation}
\frac{\partial n_{m}}{\partial t}=P_{m}-\left( \gamma _{R}+R|\psi
_{m}|^{2}\right) n_{m},~m=1,2.  \label{rate}
\end{equation}%
\begin{figure}[tbp]
\centering
\includegraphics[width=0.8\linewidth]{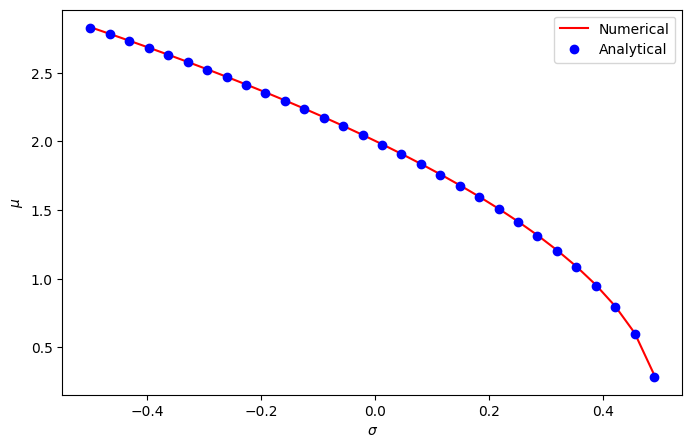}
\caption{Chemical potential $\mu$
of VAV states as a function of the SOC strength $\sigma$, for
fixed norm $N = 1$ satisfying Eq. (\ref{N}).
The analytical expression given by Eq. (\ref{eigen}), with the nonlinearity-induced correction (\ref{mu1}), 
is compared to the numerical results produced by solving the eigenvalue problem (\ref{uVAV}) with
$\delta g = 0.01$ and in-plane trapping frequency $\omega = 1$. } 
\label{mu_vs_sig}
\end{figure}
In this connection, the terms in Eq. (\ref{gpe}) proportional to $%
n_{3-m},n_{m}$ and $\gamma _{c}$ represent the cumulative effects of the
incoherent pump [whose rate is represented by term $P_{m}$ in Eq. (\ref{rate}%
)] and losses, with $R$ being the rate of scattering of excitons from the
reservoir density into the condensate, while $\gamma _{c}$ is the polariton
decay rate. A crucial condition is that the excitonic reservoir density $%
n_{m}$ reaches a steady-state value, at which the reservoir pump ($Rn_{m}$)
exactly compensates the decay rate $\gamma _{c}$ \cite{zezyulin2018spin}. We
focus the study on the regime in which the EP-EP interaction, determined by
factor $g(|\psi _{1}|^{2}+|\psi _{2}|^{2}$), dominates over the reservoir
interaction ($g_{R}n_{m}$, $\bar{g_{R}}n_{3-m}$). Thus, the mutually
compensated pump and loss terms may be neglected, and the condensate
evolution is governed by the remaining nonlinear and SOC terms, effectively
mimicking the conservative system \cite{sakaguchi2017spin}. In this case,
Eqs. (\ref{gpe}) simplify to
\begin{figure}[tbp]
\centering
\includegraphics[width=0.8\linewidth]{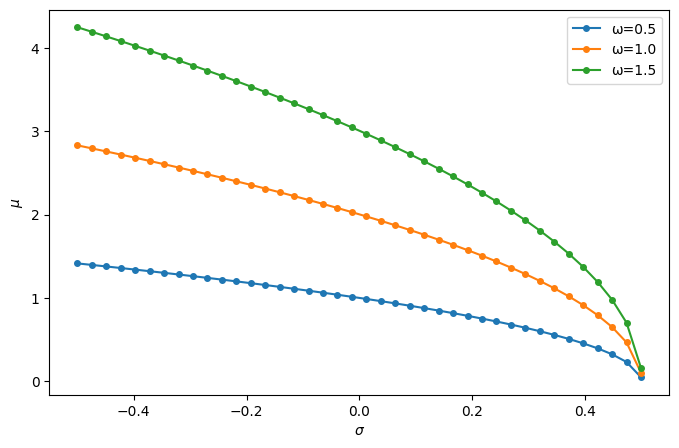}
\caption{Chemical potential $\mu$ of VAV states as a function of the SOC strength $\sigma$, 
for distinct values of the in-plane trapping frequency $\omega$, as given by analytical expression Eq. (\ref{eigen}), with the nonlinearity-induced correction (\ref{mu1}), taken for $\delta g =0.01$
and a fixed norm $N =1$, satisfying Eq. (\ref{N}).}
\label{mu_vs_sig_vary}
\end{figure}
\begin{subequations}
\label{egpe}
\begin{gather}
i\frac{\partial \psi _{1}}{\partial t}=\bigg[-\frac{1}{2}\nabla ^{2}+\frac{%
\delta g}{2}\left( |\psi _{1}|^{2}+|\psi _{2}|^{2}\right) +U(r) \nonumber\\
+g\left( |\psi _{1}|^{2}-|\psi _{2}|^{2}\right) \bigg]\psi _{1}\\
+\sigma \left( \frac{\partial }{\partial x}-i\frac{\partial }{\partial y}%
\right) ^{2}\psi _{2},\nonumber \\
i\frac{\partial \psi _{2}}{\partial t}=\bigg[-\frac{1}{2}\nabla ^{2}+\frac{%
\delta g}{2}\left( |\psi _{2}|^{2}+|\psi _{1}|^{2}\right) +U(r)\nonumber \\
-g\left( |\psi _{2}|^{2}-|\psi _{1}|^{2}\right) \bigg]\psi _{2} \\
+\sigma \left( \frac{\partial }{\partial x}+i\frac{\partial }{\partial y}%
\right) ^{2}\psi _{2}\nonumber.
\end{gather}
\end{subequations}
\newline
The scaled form of the conservative GP equations (\ref{egpe}) uses units for
lengths, time, energy, and density ($|\psi |^{2}$) defined, respectively, as
$l=\sqrt{\hbar /m\omega _{\perp }}$, $\omega_{\perp} ^{-1}$, $\hbar \omega _{\perp }$, $l^{-2}$.

\section{Numerical results}

Following Ref. \cite{Ben-Li}, it is convenient to rewrite the 2D equations (%
\ref{egpe}) in the polar coordinates, $\left( r,\theta \right) $, with the
linear SOC operators expressed by means of the identity
\begin{equation}
\frac{\partial }{\partial x}\pm i\frac{\partial }{\partial y}\equiv \exp
\left( \pm i\theta \right) \left( \frac{\partial }{\partial r}\pm \frac{i}{r}%
\frac{\partial }{\partial \theta }\right) .  \label{operator}
\end{equation}%
Stationary solutions to Eq. (\ref{egpe}), with chemical potential $\mu $,
are sought in the conventional form.
\begin{equation}
\psi _{1,2}=\exp \left( -i\mu t+iS_{1,2}\theta \right) u_{1,2}(r),
\label{mu}
\end{equation}%
where $S_{1,2}$ are integer vorticities, which are related by equality%

\begin{figure*}[tbp]
\centering
\xincludegraphics[width=0.6\textwidth, label=\textbf{}]{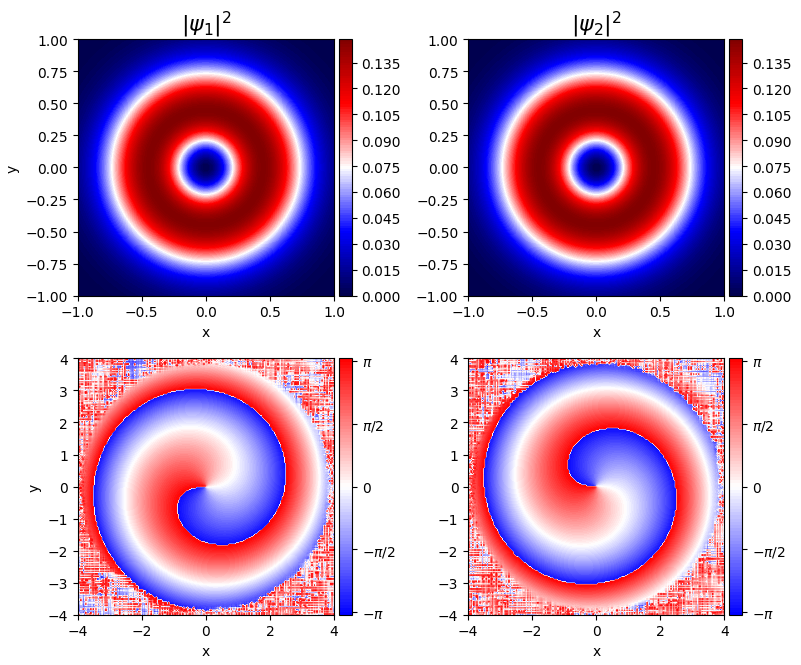}\
\caption{A stable VAV bound state, obtained as a stationary solution of Eqs. (\protect\ref{egpe}) with coefficients $\protect\delta g=0.1$, $g=1$ and $%
\protect\sigma =0.48$. Contour plot of densities $|\protect\psi _{1}{x,y}%
|^{2}$ and $|\protect\psi _{2}(x,y)|^{2}$ are shown for frequencies $f=33$
Hz\ and $f_{\perp }=167$ Hz, with the corresponding scale value of $\omega=33/167$ from Eq. (1), 
with chemical potential $\mu=0.301$ and norm $N=1$, see Eqs. (\ref{uVAV}) and (\ref{N}). The lower panel displays the corresponding phase profiles. In all figures, dimensionless quantities are plotted, where lengths, time, energy, and density are respectively scaled as \( \sqrt{\hbar / (m \omega_\perp)} \), \( \omega_{\perp}^{-1} \), \( \hbar \omega_{\perp} \), and \( l^{-2} \).
}
\label{fig_1}
\end{figure*}
\begin{equation}
S_{2}=S_{1}+2,  \label{12}
\end{equation}%
pursuant to the above-mentioned fact that the SOC terms are represented by
the second-order operator in Eqs.\ (\ref{egpe}). The substitution of ansatz (%
\ref{mu}) and potential (\ref{U}) in Eqs. (\ref{egpe}) gives rise to the
radial equations for real stationary wave functions $u_{1,2}(r)$:
\begin{eqnarray}
\mu u_{1} &=&-\frac{1}{2}\left( \frac{d^{2}u_{1}}{dr^{2}}+\frac{1}{r}\frac{%
du_{1}}{dr}-\frac{S_{1}^{2}}{r^{2}}u_{1}\right) +\bigg[\frac{\delta g}{2}%
\bigg(u_{1}^{2}+u_{2}^{2}\bigg)  \notag \\
&&+g\left( u_{1}^{2}-u_{2}^{2}\bigg)+\frac{\omega ^{2}}{2}r^{2}\right]
u_{1}+\sigma \bigg(\frac{d^{2}u_{2}}{dr^{2}}  \notag \\
&&+\frac{2\left( S_{1}+2\right) -1}{r}\frac{du_{2}}{dr}+\frac{S_{1}\left(
S_{1}+2\right) }{r^{2}}u_{2}\bigg),  \label{u1}
\end{eqnarray}%
\begin{eqnarray}
\mu u_{2} &=&-\frac{1}{2}\left( \frac{d^{2}u_{2}}{dr^{2}}+\frac{1}{r}\frac{%
du_{2}}{dr}-\frac{\left( S_{1}+2\right) ^{2}}{r^{2}}u_{1}\right) +\bigg[%
\frac{\delta g}{2}\big( u_{1}^{2}\notag \\
&&+u_{2}^{2}\big)+g\left( u_{1}^{2}-u_{2}^{2}\right) +\frac{\omega ^{2}}{2}r^{2}\bigg]%
u_{2}  \\
&&+\sigma \bigg(\frac{d^{2}u_{1}}{dr^{2}}-\frac{2S_{1}+1}{r}\frac{du_{1}}{%
dr}+\frac{S_{1}\left( S_{1}+2\right) }{r^{2}}u_{1}\bigg) \notag.
\label{u2}
\end{eqnarray}
Most fundamental solutions admitted by relation (\ref{12}) represent VAV
states, with
\begin{equation}
S_{1}=-1,~S_{2}=+1,  \label{simplest}
\end{equation}%
and SV ones, with
\begin{equation}
S_{1}=0,~S_{2}=2.  \label{SV}
\end{equation}%
\begin{figure*}[th]
\centering
\includegraphics[width=1.0\linewidth]{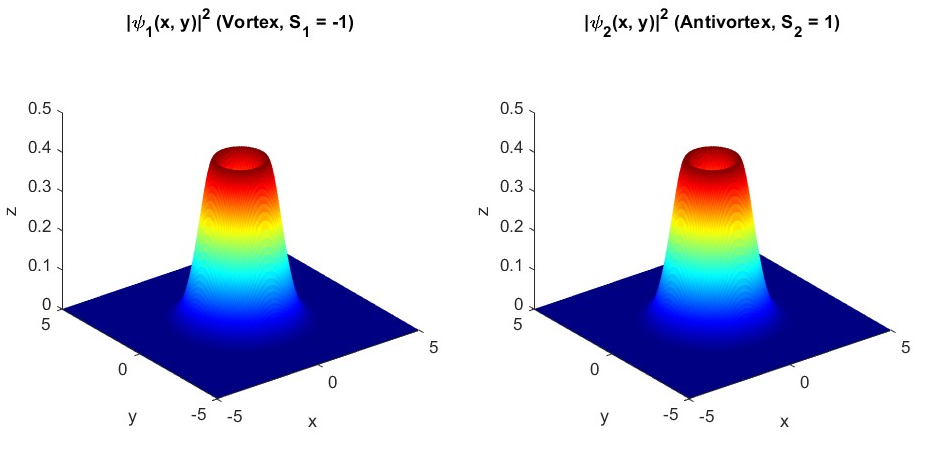}
\caption{Three-dimensional profiles of the VAV bound state from Fig. \protect
\ref{fig_1}, with vorticities $S=-1$ and $S=1$ of the two components for chemical potential $\mu=0.301$ and norm $N=1$ satisfying Eq. (\ref{uVAV}) and Eq. (\ref{N}).}
\label{Time_ev}
\end{figure*}
We here focus on the VAV state (\ref{simplest}), as it has the best chances
to be stable \cite{sakaguchi2017spin}. For this solution, with $%
u_{1}(r)=u_{2}(r)\equiv u(r)$, Eqs. (\ref{u1}) and (\ref{u2}) reduce to the
single radial equation:%
\begin{eqnarray}
\mu u&=&\left( -\frac{1}{2}+\sigma \right) \left( \frac{d^{2}u}{dr^{2}}+\frac{1%
}{r}\frac{du}{dr}-\frac{1}{r^{2}}u\right) +\delta g\cdot u^{3}\nonumber\\
&&+\frac{\omega^{2}}{2}r^{2}u. 
\label{uVAV}
\end{eqnarray}%
We initially discuss the linear system by setting $\delta g=0$. In this
case, the exact solution to Eq. (\ref{uVAV})
amounts to the commonly known 2D HO wave function,
\begin{equation}
u(r)=u_{0}r\exp \left( -\frac{\omega }{2\sqrt{1-2\sigma }}r^{2}\right) ,
\label{u0}
\end{equation}%
with the respective eigenvalue of the chemical potential,%
\begin{equation}
\mu =2\sqrt{1-2\sigma }\omega .  \label{eigen}
\end{equation}%
Here, $u_{0}$ is an arbitrary amplitude, and this solution exists in the
case of $\sigma <1/2$ (including all values of $\sigma <0$). Considering the nonlinear term in Eq. (\ref{uVAV}) as a perturbation, the first-order correction to the chemical potential is produced by the straightforward application of the quantum-mechanical perturbation
theory to this equation:
\begin{equation}
\mu ^{(1)}=\frac{u_{0}^{2}(g+g_{12})}{4}\sqrt{1-2\sigma },  \label{mu1}
\end{equation}%
where expression (\ref{delta-g}) is substituted for $\delta g$. The corresponding bound states are characterized by the
normalized number of particles (norm),
\begin{equation}
N=\int_{-\infty }^{+\infty }\left( |\psi _{1}|^{2}+|\psi _{2}|^{2}\right)
dxdy\equiv N_{1}+N_{2},  \label{N}
\end{equation}%
which is a dynamical invariant of the system.

The mean-field equations (\ref{gpe}) were solved numerically. Usually, the
split-step Fourier-transform method \cite%
{javanainen2006symbolic,lakoba2020study,semenova2021comparison}, the
Cranck-Nicolson one \cite{young2023openmp}, and the pseudo-spectral method
\cite{antoine2021scalar} are employed to produce numerical solutions. We
here used the split-step algorithm, presenting the results in a normalized
form. For the formation of VAV bound states, we fix realistic values of the
parameters, \textit{viz}., the SOC strength $\sigma =0.48$ \cite{sakaguchi2017spin} trapping frequency $f=33$ Hz, and the transverse one $f_{\perp }=167$ Hz. The results also depend on the interaction coefficients, which are determined by the scattering length of atomic collisions and norm Eq. (\ref{N}) loaded into the HO trap. Thus, numerical solutions for VAV bound states were obtained by solving Eq. (\ref{egpe}) in real time, with respective
initial states.
For instance, to produce a $2\times 2$ square VAV\ lattice, the initial states should be of the same type. The stability is possible as a result of the interplay of the confinement imposed by the holding potential and repulsive interaction between the VAV pixels. The vortices of the same sign, with either $S = +1$ or $S=-1$, repel each other in each component. The repulsive pairwise interactions between the pixels are an obvious stabilizing factor for multi-pixel patterns.

To demonstrate the formation of a single VAV bound state in the HO potential
trap, we take interaction coefficients $g=1$, $\delta g=0.1$.
In this case, the VAV bound state exhibits the stable vortex pattern
displayed in Fig. \ref{fig_1} by means of 2D contour density plot of $|\psi
(x,y)|^{2}$ for each component,
in the $\left( x,y\right) $ plane.
The lower panels of the
figure represent the corresponding phase structures, \textit{viz}.,
spirals winding around the vortex pivot. Further, the 3D profiles of the
VAV bound state from Fig. \ref{fig_1} are plotted in Fig. \ref{Time_ev}.
The VAV states, characterized by dependencies of their chemical potential on the SOC strength $\sigma$, which are plotted in Fig. \ref{mu_vs_sig}, 
showing the comparison between the analytical and numerical findings. The analytical expression (\ref{eigen}), with the nonlinearity-induced correction (\ref{mu1}), is consistent with the numerical results, obtained by solving Eq. (\ref{uVAV}) for a fixed norm $N=1$. Here we consider small $\delta g$, with the self-repulsion and cross-attraction nearly canceling each other through the Feshbach resonance. Recent work \cite{navadeh2019polaritonic} have demonstrated similar tuning techniques in exciton-polariton systems. Figure \ref{mu_vs_sig_vary} displays the chemical potential as a function of SOC strength $\sigma$, for distinct values of in-plane trapping frequency $\omega$, using the analytical expression Eq. (\ref{eigen}) with the nonlinearity-induced correction Eq. (\ref{mu1}) for a fixed norm N=1.

\begin{figure}[tph]
\centering
\xincludegraphics[width=1\linewidth,label=\textbf{a)}]{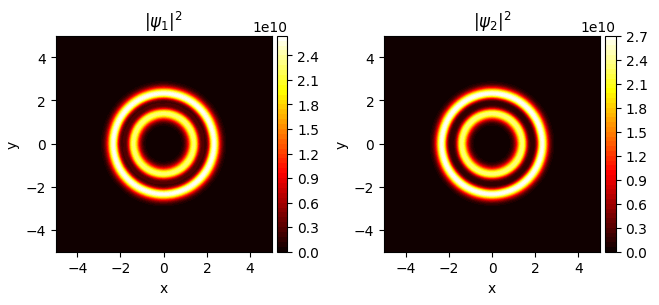}\ %
\xincludegraphics[width=1\linewidth,label=\textbf{b)}]{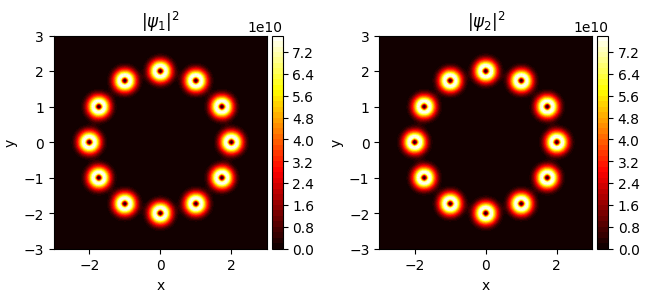}\ %
\xincludegraphics[width=1.01\linewidth,label=\textbf{c)}]{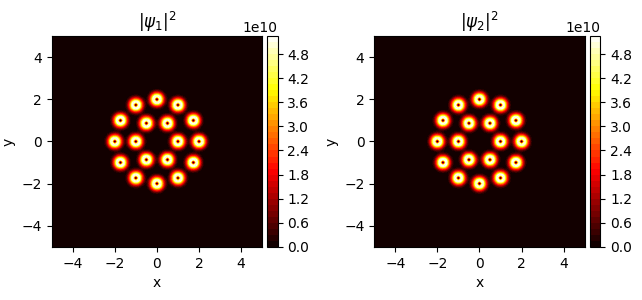}\ %
\xincludegraphics[width=1\linewidth,label=\textbf{d)}]{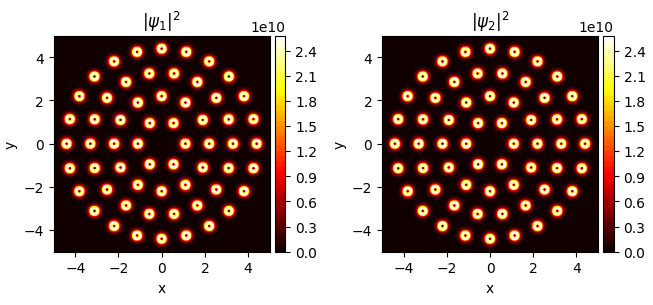}\ %
\xincludegraphics[width=1\linewidth, label=\textbf{e)}]{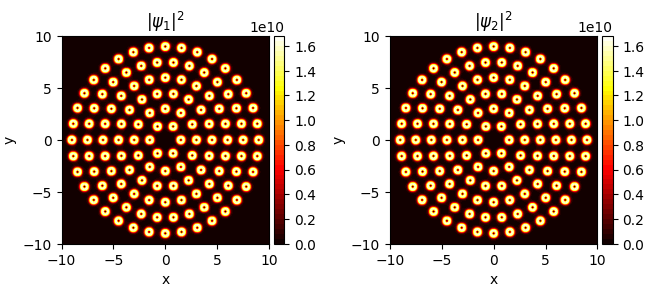}
\caption{Density distributions of $|\protect\psi _{1}(x,y)|^{2}\ $and $|%
\protect\psi _{2}(x,y)|^{2}$ for different values of the interaction
parameters, under the action of the in-plane and transverse trapping
frequencies $f=33$ Hz and $f_{\perp }=167$ Hz, respectively, corresponding to a scaled frequency of $\omega=33/167$. The plots
exhibit stable ring-shaped and annular patterns composed of multiple VAV
elements (\textquotedblleft pixels"). In a) $g=350$, $\protect\delta g=250$;
in b) $g=50$, $\protect\delta g=10$; in c) $g=100$, $\protect\delta g=50$;
in d) $g=200$, $\protect\delta g=100$; in e), $g=300$, $\protect\delta g=200$%
. In all cases, the SOC strength is $\protect\sigma =0.48$.}
\label{2}
\end{figure}

To create various lattice patterns built of multiple localized VAV elements
(\textquotedblleft pixels"), nonlinearity parameters $g$ and $\delta g$
should be selected appropriately in Eq. (\ref{egpe}), the initial state
being selected as ring, circular, square, rectangle, hexagonal and
triangular lattices. The study of these patterns begins with the ring and
circular arrangements, represented in the polar coordinates. The trap is
characterized by the in-plane and transverse trapping frequencies $f=33$ Hz
and $f_{\perp }=167$ Hz, respectively, implying a less tight in-plane trap,
that can accommodate a substantial number of particles and promote the
formation of a circular pattern. As the interaction parameters $g$ and $%
\delta g$ increase, the trap accumulates a greater number of particles, thus
leading to the generation of an increasing number of individual VAV
elements. To investigate the formation of a circular pattern, we take $g=350$
and $\delta g=250$, which produces a double-ring-shaped VAV mode, as shown
in Fig. \ref{2}a) by means of contour plot of densities $|\psi
_{1}(x,y)|^{2} $ and $|\psi _{2}(x,y)|^{2}$ in the $\left( x,y\right) $
plane. By augmenting the spacing between individual VAV pixels and
maintaining $g=50$ with $\delta g=10$, the transition of the ring-shaped
vortex to the circular array built of $12$ pixels is observed in Fig. \ref{2}%
b). Note that the creation of the latter configuration makes it necessary to
use larger interaction parameters ($g$ and $\delta g$). For $g=100$ and $%
\delta g=50$, an annular pattern composed of $18$ pixels, with $6$ in the
inner ring and $12$ in the outer one, is shown in Fig. \ref{2}c). Further
increase to $g=200$, with $\delta g=100$ results in the pattern created of $%
96$ elements, including $6$ ones in the first ring, and $12$, $18$, and $24$
elements in the second, third, and fourth rings, as shown in Fig. \ref{2}d).
For $g=300$ and $\delta g=200$, the pattern includes $6$, $12$, $18$, and $%
24 $ pixels in the first, second, third, and fourth rings, in addition to $36$
in the fifth one, as plotted in Fig. \ref{2}e) (for $g<300$ and $\delta g$ $%
<200$, the annular pattern does not exhibit the fifth stable ring). In Figs.
(\ref{2}b) to (\ref{2}e), we observe that the total number of VAV elements
in each ring is $6n$, where $n$ is the ring's number. Furthermore, in Figs. (%
\ref{2}c) to (\ref{2}e), radii of the succeeding rings grow by $1$, while
the position at the origin remains empty in all the patterns displayed in
Fig. \ref{2}.
\begin{figure}[tph]
\centering
\xincludegraphics[width=1\linewidth,label=\textbf{a)}]{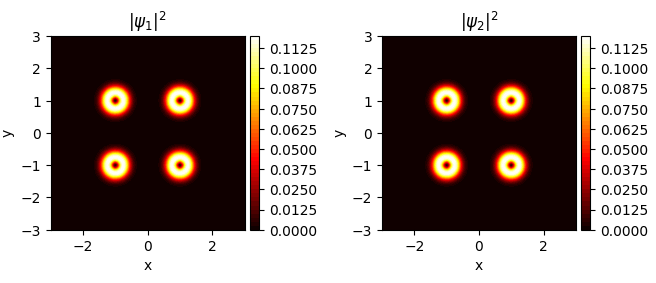}\ %
\xincludegraphics[width=1\linewidth,label=\textbf{b)}]{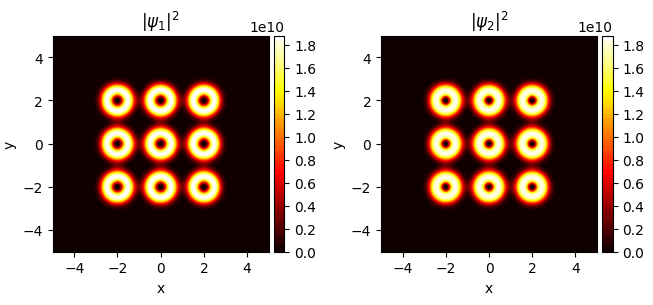}\ %
\xincludegraphics[width=1\linewidth,label=\textbf{c)}]{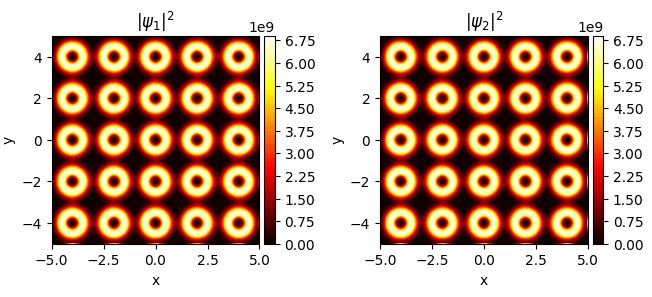}\ %
\xincludegraphics[width=1\linewidth,label=\textbf{d)}]{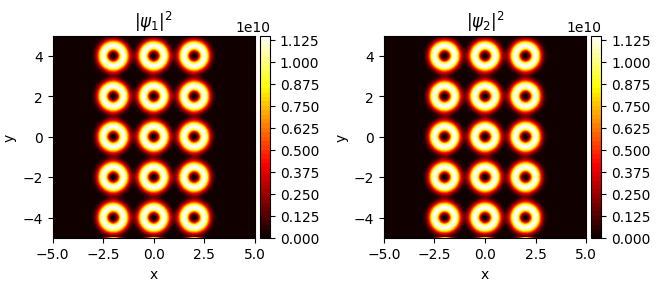}\ %
\xincludegraphics[width=1\linewidth,label=\textbf{e)}]{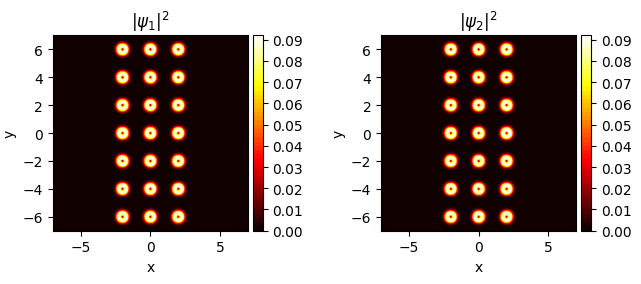}\
\caption{Density distributions $|\protect\psi _{1}{x,y}|^{2}|$ and $|\protect%
\psi _{2}(x,y)|^{2}$ are plotted by means of the contour plots for different
values of the nonlinearity parameters, with SOC strength $\protect\sigma %
=0.48$ and in-plane and transverse trapping frequencies $f=33~$Hz and $%
f_{\perp }=167$ Hz, resulting in a scaled frequency of $\omega=33/167$. a) The stable $2\times 2$ configuration
for $g=1$, $\protect\delta g=0.1$. b) The stable $3\times 3$ square lattice
for $g=2$ and $\protect\delta g=1$. c) The stable $5\times 5$ lattice, for $%
g=20$, $\protect\delta g=2$. d) The stable $3\times 5$ rectangular lattice,
for $g=10$, $\protect\delta g=0.1$ e) The stable $3\times 7$ rectangular
lattice, for $g=20$, $\protect\delta g=0.1$.}
\label{4}
\end{figure}

Next, we address square- and rectangular-lattice configurations of size $%
m\times n$ maintained by the HO trap. Some of these configurations, were obtained for low values of interaction coefficients. In comparison to the circular lattices displayed in Fig. \ref{2}, much
smaller values of $g$ and $\delta g$ are sufficient to create square-lattice
patterns. Figure (\ref{4}a) displays contour plots of 2D density profiles in
the $\left( x,y\right) $ plane for $g=1$ and $\delta g=0.1$, with the same
SOC strength as fixed above, $\sigma =0.48$. Under these conditions, in Fig. %
\ref{4}a) the system produces a $2\times 2$ square-lattice structure with
the empty position at the origin. For $g=2$ and and $\delta g=1$, a $3\times
3$ square lattice, built of $9$ VAV\ pixels (with one occupying the central
position) is observed in Fig. \ref{4}b). Next, the $5\times 5$ lattice state
is displayed in Fig. \ref{4}c), for $g=20$ and $\delta g=2$. It is composed
of $25$ pixels, including one located at the center.

\begin{figure}[tph]
\centering
\xincludegraphics[width=0.9\linewidth,label=\textbf{a)}]{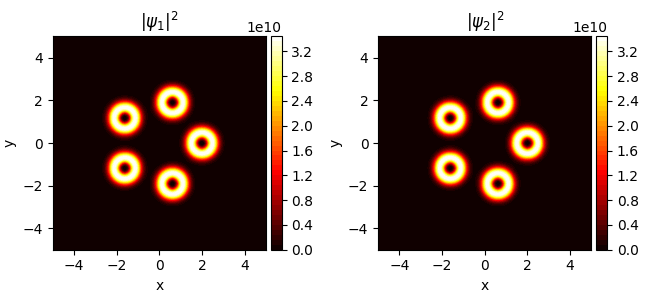}\ %
\xincludegraphics[width=0.9\linewidth,label=\textbf{b)}]{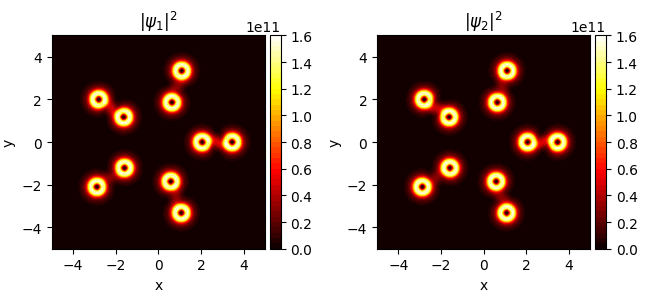}\ %
\xincludegraphics[width=0.9\linewidth,label=\textbf{c)}]{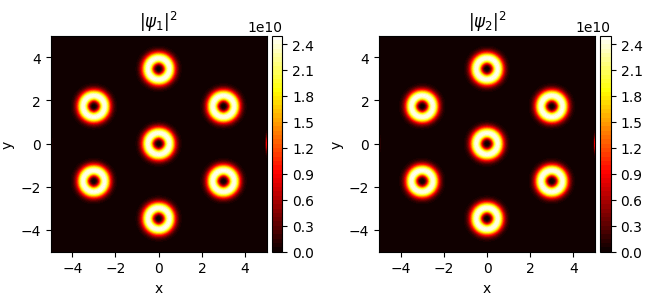}\ %
\xincludegraphics[width=0.9\linewidth,label=\textbf{d)}]{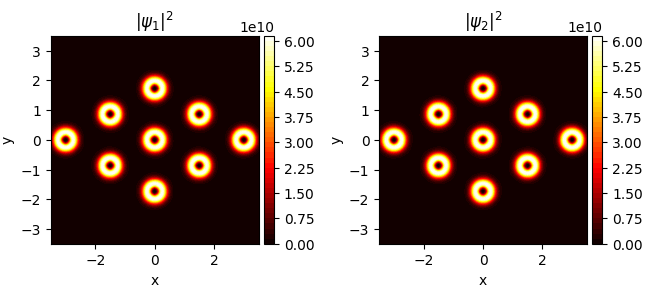}\ %
\xincludegraphics[width=0.9\linewidth,label=\textbf{e)}]{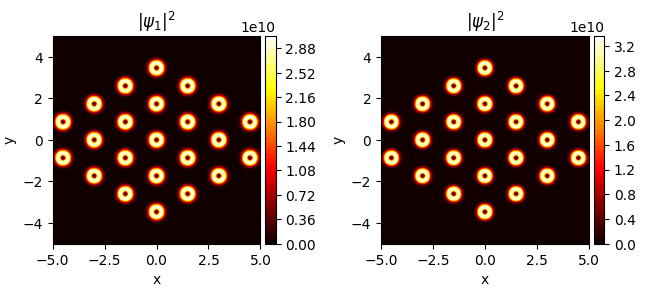}\ %
\xincludegraphics[width=0.9\linewidth,label=\textbf{f)}]{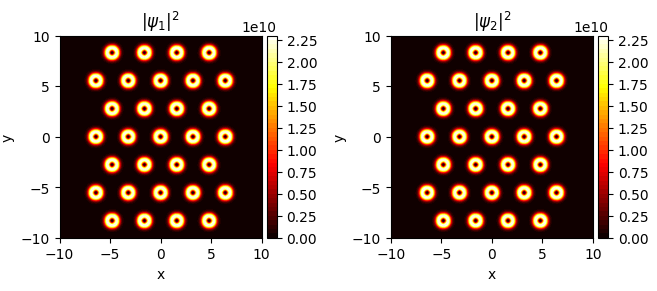}
\caption{Density contour plots of $|\protect\psi _{1}{x,y}|^{2}$ and $|%
\protect\psi _{2}(x,y)|^{2}$, produced by the numerical solution of Eqs. (%
\protect\ref{egpe}) under the action of the in-plane and transverse HO
potentials $f=33$ Hz and $f_{\perp }=167$ Hz,
respectively,  with an scaled frequency of $\omega = 33/167$ and SOC strength $\protect\sigma =0.48$. Displayed are
pentagonal chains in a) and b); the hexagonal chain in c); and diverse triangular
patterns in d) - f). The respective nonlinearity coefficients are: a) $g=50$%
, $\protect\delta g=10$; b) $g=150$, $\protect\delta g=100$ c) $g=100$, $%
\protect\delta g=50$; d) $g=450$, $\protect\delta g=250$; e) $g=650$, $%
\protect\delta g=350$; f) $g=1000$, $\protect\delta g=500$.}
\label{5}
\end{figure}

Stable rectangular lattices composed of $5\times 3$ and $7\times 3$ VAV
elements are displayed in Figs. (\ref{4}d) and (\ref{4}e). The respective
values of the nonlinearity parameters are $g=10$, $\delta g=0.1$ and $g=20$,
$\delta g=0.1$, respectively. Finally, we demonstrate stable pentagonal,
hexagonal, and triangular lattices composed of VAV pixels. Actually, the
hexagonal pattern may be considered as a triangular lattice with vacant
central positions. Therefore, with inappropriate initial condition or
parameter values, the numerical calculations may instead converge to
circular or triangular states, by filling the vacant positions. For $g=50$
and $\delta g=10$, the pentagonal ring-shaped chain, consisting of five VAV
elements, is depicted in Fig. \ref{5}a). The double pentagonal chain, plotted in Fig. \ref{5}b), corresponds to $g=150$ and $%
\delta g=100$. For $g=100$ and $\delta g=50$, the body-centered structure
displayed in Fig. \ref{5}c) exhibits six elements forming a hexagonal cell,
and one placed at the central location. In Fig. \ref{5}d), the hexagonal
structure is transformed into a triangle configuration (which also seems as
a rhombus) by adding two VAV elements at lateral positions (along the $x$
axis), in the case of $g=450$ and $\delta g=250$, so that the total number
of the elements (pixels) in the configuration is nine. In Fig. \ref{5}e),
the triangular lattice is further expanded to comprise $23$ elements, at $%
g=650$ and $\delta g=350$. Finally, Fig. \ref{5}f) demonstrates that the
system with very large nonlinearity coefficients, such as $g=1000$ and $%
\delta g=500$, maintains very large perfect triangular lattices. In
particular, the one shown in Fig. \ref{5}f) is composed of $31$ VAV pixels
arranged in four rows of $4$ pixels and three rows of $5$ ones.

\begin{figure}[tph]
\centering
\xincludegraphics[width=0.9\linewidth,label=\textbf{a)}]{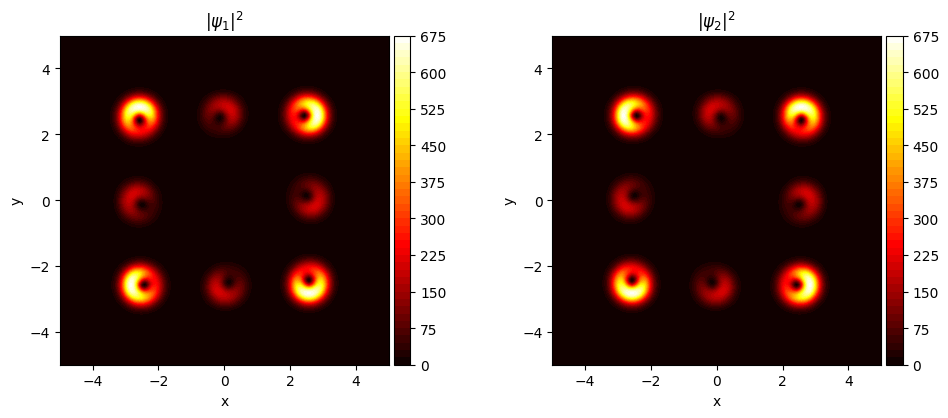}\ \xincludegraphics[width=0.9\linewidth,label=\textbf{b)}]{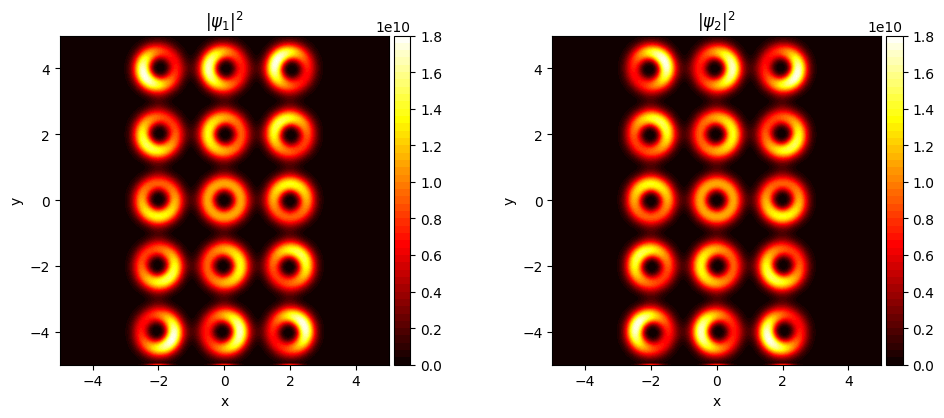}\
\xincludegraphics[width=0.9\linewidth,label=\textbf{c)}]{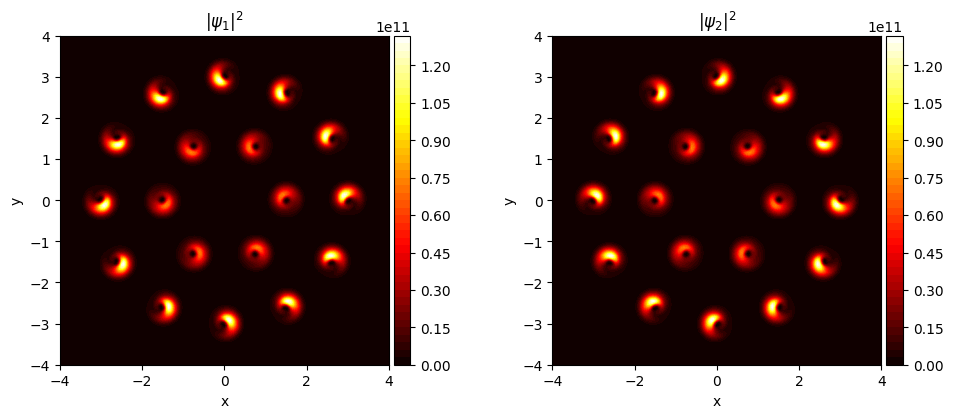}\
\caption{Density contour plots of $|\protect\psi _{1}{x,y}|^{2}$ and $|%
\protect\psi _{2}(x,y)|^{2}$, produced by the numerical solution of Eqs. (%
\protect\ref{egpe}) under the action of the strong harmonic confinement, with $f\gg f_{\perp}$ and SOC strength $\protect\sigma =0.48$. Displayed are square-shaped and rectangular states in a) and b), and the circular configuration in c). a) A $3\times3$ square configuration for $g=1$ and $\delta g=0.1$ with $f=33$ Hz, $f_{\perp}=1$ Hz, such that $\omega=33$. b) A $3\times5$ rectangular  configuration for $g=10$ and $\delta g=0.1$ with $f=33$ Hz, $f_{\perp}=10$ Hz (hence, $\omega=33/10$). c) A circular chain of VAV pixels for $g=100$ and $\delta g=50$ with $f=10$ Hz, $f_{\perp}=1$ Hz, the corresponding scaled frequency being $\omega=10$.}
\label{8}
\end{figure}

Now, the investigation is carried out for the VAV pixels under strong isotropic confinement, with $f\gg f_{\perp}$, in square-shaped, rectangular, and circular configurations. We initially discuss a \(3 \times 3\) square lattice configuration under the strong confinement, with an in-plane trapping frequency \( f = 33 \) Hz and a transverse trapping frequency \( f_{\perp} = 1 \) Hz. The interaction coefficients are set as \( g = 1 \) and \( \delta g = 0.1 \), as illustrated in Fig. \ref{8}a). We observe that the square-shaped $3\times3$ lattice is not fully compatible with the isotropic trap, whereas the simpler $2\times2$ configuration is. When subjected to the stronger isotropic confinement, the system eliminates the interstitial pixels, transforming the $3\times3$ lattice into a $2\times2$ configuration that better aligns with the isotropic trap. Figure \ref{8}b) illustrates a rectangular lattice configuration under in-plane and transverse trapping frequencies $f=33$ Hz and $f_{\perp}=10$ Hz. While the lattice is initiated as a $5\times3$ structure with the central site, the limited available space prevents the central site from fully developing, and some sites also fail to exhibit a complete vortex profile.
Finally, a circular vortex configuration represented in Fig. \ref{8}c) is investigated under the confinement with $f=10$ Hz and $f_{\perp}=1$ Hz, with interaction parameters $g=100$ and $\delta g=50$. As expected, the strong confinement makes the inner vortex ring less visible, due to insufficient space for its formation. As the confinement strengthens, the density in the trap significantly increases, restricting the ability of the vortices to rearrange freely. Additionally, the high density suppresses the formation of well-defined vortices, as the limited space restricts their full development.

\section{Conclusion}
In the framework of the 2D mean-field model of EP (exciton-polariton)
condensates, we have performed the analysis of the formation of the single
VAV (vortex-antivortex)\ two-component bound state, and lattices\ build of
several or many VAV elements (\textquotedblleft pixels") under the action of
the photonic SOC (spin-orbit coupling), represented by the second-order
differential operator , and in-plane HO (harmonic-oscillator) trapping
potential. The model is considered in the conservative form under
the condition of the compensation of losses and pump. Making use of the
trap's capacity to accommodate a substantial number of particles, the
examination of various lattice patterns was conducted, utilizing appropriate
values of the interaction strengths $g$ and $\delta g$ and other parameters.
The study of various VAV lattices begins with circular patterns composed of
two concentric rings. Further, circular lattices built of VAV elements were
found as single- and multi-layer patterns, with the number of elements in
each circular layer being a multiple of $6$. Then, square and rectangular
lattices, composed of the same VAV elements, were investigated. The lowest $%
2\times 2$ lattice state, with the unoccupied central position, was
constructed for $g=1$ and $\delta g=0.1$, while the $3\times 3$ lattice,
composed of $9$ VAV pixels, was constructed for $g=2$ and $\delta g=1$,
exhibiting the occupied central position. Stable rectangular lattices of
sizes $3\times 5$ and $3\times 7$ were found for $g=10$, $\delta g=0.1$ and $%
g=20$, $\delta g=0.1$, respectively. The formation of pentagonal, hexagonal
and triangular VAV lattices was addressed too. Triangular lattices were
found at large values of the interaction parameters. A stable pentagonal
ring was produced for $g=50$ and $\delta g=10$, while a double pentagonal
ring employs $g=150$ and $\delta g=100$. The hexagonal lattice was produced
for the respective minimum values of the coefficients, $g=100$ and $\delta
g=50$, composed of six VAV pixels forming the hexagonal cell, with an extra
pixel placed at the center. To construct the triangular lattice state
composed of nine elements, by adding two elements to the hexagonal cell, $g$
and $\delta g$ had to be set to large values, \textit{viz}., $450$ and $250$%
, respectively. To add an extra layer to the triangular lattice, making the
total number of constituent VAV pixels equal to $23$, the interaction
coefficients were further ramped up to $g=650$ and $\delta g=350$. With very
large values of the nonlinearity coefficients, $g=1000$ and $\delta g=500$,
a large stable triangular lattice, composed of $31$ elements, was found. The
odd number of constituents takes place in those patterns which include a VAV
pixel occupying the center's position.
The formation of the various lattice patterns in our system
is determined by the interaction coefficients and initial conditions, with a
critical density required for the creation of stable patterns being proportional to
the interaction coefficients. Different configurations require disparate
interaction levels: in the circular patterns, the lattice keeps the center
empty, leading to a more uniform density distribution, whereas in
hexagonal and triangular configurations the center is occupied, necessitating to use stronger interactions to reach the critical density
for the pattern formation. It is remarkable that the various lattice
patterns, which were previously reported in free-space models, persist under
the action of the weak HO trap. They exhibit dynamical robustness and
suggest additional theoretical and experimental studies. Further, we investigated the behavior of VAV pixels under 
the action of the strong isotropic confinement in different geometries, including square-shaped, rectangular, and circular arrangements. Our findings reveal that a $3 \times 3$ square lattice is not fully compatible with the isotropic trapping for the in-plane and transverse trapping frequencies $f=33$ Hz and $f_{\perp}=1$ Hz, as stronger confinement leads to the elimination of interstitial pixels, resulting in a more stable \(2 \times 2\) configuration. Similarly, for a rectangular lattice with $f=33$ Hz and $f_{\perp}=10$ Hz, spatial constraints prevent the full development of certain vortex sites, particularly in the central region. Finally, for a circular vortex configuration, the strong confinement with $f=10$ Hz and $f_{\perp}=1$ Hz suppresses the inner vortex ring, restricting the free rearrangement of vortices due to increased density in the trap.
The stability of these patterns should permit access to them in the experiment by choosing appropriate initial conditions in EP systems featuring the underlying TE-TM splitting.

The analysis may be enhanced by incorporating the Zeeman splitting between
the two components of the spinor wave function \cite%
{nalitov2015polariton,karzig2015topological}. It would be interesting to further explore the interaction between two VAV pixels with opposite polarities, i.e., one with vorticities $(+1,-1)$ and the other one with $(-1,+1)$. Since these configurations are expected to interact attractively, they may form bound states such as mutually orbiting vortex pairs. Future work may focus on analyzing such interactions in greater detail to understand the formation of more complex vortex structures. Further extension may include the examination of lattice patterns composed of VAV pixels in dissipative
SOC systems, where a variety of spatially periodic states may be expected
\cite{vercesi2023phase,chen2024nonequilibrium}. \newline
\newline
\textbf{Acknowledgment:} S. Sanjay and S. Saravana Veni acknowledge Amrita
Vishwa Vidyapeetham, Coimbatore, where this work was supported under Amrita
Seed Grant (File Number: ASG2022141). The work of B. A. Malomed was
supported, in part, by the Israel Science Foundation through grant No.
1695/22.


\end{document}